\documentstyle[aps,prl,twocolumn]{revtex}
\begin{document}
\draft
\def\B.#1{{\bbox{#1}}}
\title{Fusion Rules and Conditional Statistics in Turbulent Advection} 
\author {Emily S.C. Ching$^*$, Victor S. L'vov$^{\dag}$
 and Itamar Procaccia$^{\dag}$ }
\address{$^*$ Department of
Physics, the Chinese University of Hong Kong, Shatin, Hong Kong \\
and $^{\dag} $Department of Chemical Physics,
The Weizmann Institute of Science, Rehovot 76100, Israel } 
\maketitle 
\begin{abstract}
%%%%%%%%%%%%%%%%%%%%%%%%%%%%%%%%%%%%%%%%%%%%% 1
  Fusion rules in turbulence address the asymptotic properties of
  many-point correlation functions when some of the coordinates are
  very close to each other. Here we put to experimental test some
  non-trivial consequences of the fusion rules for scalar correlations
  in turbulence. To this aim we examine passive turbulent advection as
  well as convective turbulence. Adding one assumption to the fusion
  rules one obtains a prediction for universal conditional statistics
  of gradient fields. We examine the conditional average of the scalar
  dissipation field $\left<\nabla^2
    T(\B.r)|T(\B.r+\B.R)-T(\B.r)\right>$ for $R$ in the inertial
  range, and find that it is linear in $T(\B.r+\B.R)-T(\B.r)$ with a
  fully determined proportionality constant.  The implications of
  these findings for the general scaling theory of scalar turbulence
  are discussed.
\end{abstract}
%%%%%%%%%%%%%%%%%%%%%%%%%%%%%%%%%%%%%%%%%%%%%%%%%%%%%%%%%%%%%%%%%%%%%% 
The aim of this Letter is to present new analysis of experimental data
\cite{Sre,87HCL,89SWL} pertaining to turbulent scalar advection and to
discuss the implications of this analysis in the context of fusion
rules and conditional averages. We begin with a short theoretical
background to the issues in order to make this Letter self contained.
Turbulent advection is described mathematically by the equation of
motion for a scalar field T(\B.r,t)
\begin{equation}
\big[\partial_t +
\B.u(\B.r,t) \cdot \bbox{\nabla}\big] T(\B.r,t)
 = \kappa \nabla^2 T(\B.r,t) \ , 
\label{advect} \end{equation}
where $\kappa$ is the scalar diffusivity and $\B.u(\B.r,t)$ is the
turbulent velocity field responsible for the advection of $T(\B.r,t)$.
The problem of ``passive" scalar advection is the one in which the
properties of $\B.u(\B.r,t)$ are not affected by those of the scalar
$T(\B.r,t)$. In ``active" scalar problems, like turbulent convection,
the velocity field and its statistical properties are coupled with
those of the scalar field, and Eq.  (\ref{advect}) has to be
supplemented with an additional equation for $\B.u(\B.r,t)$ and 
$T(\B.r,t)$. In our
thinking below we consider passive as well as active scalar fields. In
both cases we are interested in the limit of large Peclet number Pe
which is defined as $U_L L/\kappa$ where $U_L$ is the typical velocity
difference across the outer scale $L$ of turbulence.

The statistical properties of the scalar fields are commonly discussed
in terms of the so called ``structure functions" $S_{2n}(R)$ defined
as 
\begin{equation}
\label{s2n_def}
S_{2n}({\B.R}) \equiv
\langle [T({\B.r}+{\B.R}, t) -
T({\B.r}, t)]^{2n} \rangle \ ,
\end{equation}
where $\left<\dots\right>$ stands for an ensemble average. In writing
this equation, we assume that the statistics of the velocity field
leads to a stationary and spatially homogeneous ensemble of the scalar
$T$. If the statistics is also isotropic, then $S_{2n}$ becomes a
function of $R$ only, independent of the direction of ${\bbox R}$. The
scaling exponents of the structure functions $S_{2n}(R)$ characterize
their $R$ dependence in the limit of large Pe,
\begin{equation}
\label{zeta_def}
S_{2n}(R) \propto R^{\zeta_{2n}} ,
\end{equation}
when $R$ is in the ``inertial" interval of scales that will be
discussed later in this paper.  One of the fundamental questions in
the theory of turbulent advection is what the numerical values of
the exponents $\zeta_{2n}$ are, and whether they conform with classical
Kolmogorov type arguments, or rather exhibit the phenomenon of
multiscaling.

An important equation to analyze in this context is the so-called
``balance equation" which is obtained by writing Eq.(\ref{advect})
twice at points $\B.r$ and $\B.r+\B.R$, subtracting the equations, and
multiplying the result by $2n[T({\bbox r}+{\bbox R}, t) - T({\bbox r},
t)]^{2n-1}$. Taking the ensemble average and using the symmetry
between the two points analyzed, one finds the balance equation
\begin{equation} D_{2n}(R)=J_{2n}(R) \ , \label{bal}
\end{equation}
where $D_{2n}(R)$ stems from the convective term in
 (\ref{advect}) and $J_{2n}(R)$ stems from
the diffusion term:
\begin{equation}
J_{2n}(R)=-4n\kappa\left<\nabla^2 T(\B.r)[T({\bbox r}+{\bbox R})
 - T({\bbox r})]^{2n-1}\right>\ . \label{J2n}
\end{equation}

It was argued recently, first by Kraichnan\cite{94Kra} and later in
refs.\cite{95KYC,95FGLP,95LP-d} that balance equations play a very
important role in providing non-perturbative relations that can
determine, or severely constrain, the values of the scaling exponents
$\zeta_{2n}$. A good example is Kraichnan's model of passive scalar
advection \cite{68Kra}, in which the velocity field $\B.u$ is
delta-correlated in time, but exhibits power law scaling in space. In
this case the convective term $D_{2n}$ can be calculated exactly in
terms of $S_{2n}$ \cite{94Kra}, \begin{equation} D_{2n}(R)=
  -R^{1-d}{\partial \over \partial R} R^{d-1}h(R) {\partial \over
    \partial R} S_{2n}(R) \ , \label{D2n} \end{equation} with $d$
being the space dimension and $h(R)$ the scalar part of the
eddy-diffusivity, $h(R)\propto R^{\zeta_h}$ with $\zeta_h$ a scaling
exponent. If we could represent exactly also the RHS $J_{2n}(R)$ in
terms of $S_{2n}(R)$, we could evaluate all the scaling exponents
$\zeta_{2n}$ from the balance equation (\ref{bal}). Here is where the
fusion rules come in. The fusion rules appear naturally in the
analytic theory of Navier-Stokes turbulence
\cite{95LP-d,94LL,95LP-b,96LP} and passive-scalar turbulent advection
\cite{95FGLP,96LP,95CFKL}, and they determine the analytic structure
of $n$-point correlation functions when a group of coordinates tend
towards each other.  In the case of scalar advection we consider
simultaneous many-point correlation functions of field differences:
\begin{eqnarray}
&&{\cal F}_{2n}({\bbox r}_0|{\bbox r}_1,{\bbox r}_2\dots{\bbox r}_{2n}) 
=\label{Fn}\\ 
&&\left<\delta T({\bbox r}_0,{\bbox r}_1)\,
\delta T({\bbox r}_0,{\bbox r}_2)\dots 
\delta T({\bbox r}_0,{\bbox r}_n)\right>\ . \nonumber
\end{eqnarray}
We note that the previously defined structure functions $S_{2n}$ are
obtained by ``fusing" all the coordinates $\B.r_1\dots \B.r_{2n}$ to
one coordinate $\B.r_0+\B.R$.  The fusion rules were derived in
\cite{96LP} for systems that enjoy universality of the scaling
exponents (i.e the scaling exponents do not depend on the detailed
form of the driving of the turbulent flows), and whose correlation
functions ${\cal F}_{2n}$ are homogeneous functions of their
arguments,
\begin{equation}
{\cal F}_{2n}(\lambda{\bbox r}_0|\lambda{\bbox r}_1,\dots,
\lambda{\bbox r}_{2n})
=\lambda^{\zeta_{2n}}{\cal F}_{2n}({\bbox r}_0|{\bbox r}_1,
\dots,{\bbox r}_{2n}) \ . \label{hom}
\end{equation}
This form applies whenever all the distances
$\left|\B.r_i-\B.r_0\right|$ are in the so-called ``inertial range",
between the outer scale $L$ and the appropriate dissipative scale of
the system, denoted below as $\eta$. The fusion rules address the
asymptotic properties of ${\cal F}_{2n}$ when a group of $p$ points,
$p<2n-1$ tend towards $\B.r_0$ $(\left| \B.r_i-\B.r_0\right|\sim \rho$
for all $i\le p$), while all the other coordinates remain at a larger
distance $R$ from $\B.r_0$ $(\left|\B.r_i-\B.r_0\right|\sim R$ for
$i>p$, and $R\gg \rho$).  In particular under the two general
assumptions of scale invariance and universality of the scaling
exponents the fusion rules state that to leading order in $\rho/R$ \FL
\begin{equation}
{\cal F}_{2n}({\bbox r}_0|{\bbox r}_0+\bbox\rho,{\bbox r}_0+\B.R,
\dots,{\bbox r}_{0}+\B.R)\sim
{S_2(\rho)\over S_2(R)} S_{2n}(R). \label{fusion2} 
\end{equation}
This forms holds as long as $\rho$ is in the inertial range.

We show now how to use this fusion rule to calculate $J_{2n}(R)$.
First write it as
$$
J_{2n}(R)=-4\kappa n \lim_{\rho \to 0} \bbox\nabla^2_{\rho}{\cal
  F}_{2n}({\bbox r}_0|{\bbox r}_0+ \bbox\rho,{\bbox r}_0+\B.R,\dots,{\bbox
  r}_{0}+\B.R). $$
In using the fusion rule (\ref{fusion2}) to
evaluate this quantity we interpret the limit $\rho \to 0$ as a limit
$\rho \to \eta$. This seems natural for large Peclet numbers when
$\eta \to 0$. It is important however to stress that there is a hidden
assumption here. We expect the function ${\cal F}_{2n}({\bbox r}_0|{\bbox
  r}_0+\bbox\rho,{\bbox r}_0+\B.R,\dots,{\bbox r}_{0} +\B.R)$, which is a
function of $\bbox \rho$ and $\B.R$ to change its analytic behavior as
a function of $\rho$. This change occurs at the viscous crossover
scale $\eta$. The issue is whether this crossover scale is $n$ and $R$
independent. That this is so has been {\em proven} for Kraichnan's
model of turbulent advection \cite{96LP} but not in general. We
believe that this is more generally true due to the linearity of the
equation of motion (\ref{advect}), independently of the statistical
properties of the driving velocity field. The experimental results
which we discuss later in this Letter will strongly indicate that this
is the case in a wide context of scalar turbulent fields. We caution
the reader that this is not so in Navier-Stokes turbulence. With this
in mind we write
\begin{equation}
J_{2n}(R) \sim -4\kappa n \big[\nabla^2_{\rho}S_2(\rho)
|_{\rho=\eta}\big] S_{2n}(R)/S_2(R)\ . \
\end{equation}
Using the fact that the mean of the scalar dissipation field, denoted
$\bar\epsilon$, is evaluated as $\bar\epsilon \sim
\kappa\big[\nabla^2_{\rho}S_2(\rho)|_{\rho=\eta}\big]$, and also the
fact that in the inertial range $J_2(R)=-4\bar\epsilon$, we write
\begin{equation}
J_{2n}(R) = n C_{2n} J_2 S_{2n}(R)/S_2(R)\ , \label{almostfinal} 
\end{equation}
where $C_{2n}$ is an as yet unknown dimensionless coefficient, but
$C_2=1$. Eq.(\ref{almostfinal}) was suggested for Kraichnan's model in
\cite{94Kra} and derived in \cite{95FGLP}. Here we propose that it
holds in a much wider context. To this end we turn now to the analyzes
of experimental data.

We first display experimental results that confirm the theoretical
prediction (\ref{almostfinal}). The results show that to a good
accuracy $C_{2n}\approx 1$ for all $n$ and $R$. The theoretical
consequences of this $n$ and $R$-independence of $C_{2n}$ will be
discussed after examining the data.
%%%%%%%%%%%%%%%%%%%%%%%%%%%%%%%%%%%%%%%%%%%%%%%%%%%%%%%%%%%%%%%%%%
\begin{figure}
\caption{ A plot of $\log|J_{2n}(R)/(2n\kappa)|$ vs. 
$\log|(2\kappa)^{-1}J_2 S_{2n}(R)/ S_2(R)|$ for
$n=2$~(squares), 3~(triangles), 4~(diamonds), 5~(stars), and
6~(circles) and $R$ in the inertial range. The data are taken from
Yale [1]. The line is not a fit, but the theoretical expectation with 
slope 1  and intercept 0.}
\label{Fig1}
\end{figure}

%%%%%%%%%%%%%%%%%%%%%%%%%%%%%%%%%%%%%%%%%%%%%%%%%%%%%%%%%%%%%%%%%%
We use temperature data measured in the wake of a heated cylinder
\cite{Sre}. Water of speed 5 m/s flowed past a heated cylinder of
diameter 19 mm (Reynolds number $=9.5\times10^4$). The temperature was
measured at a fixed point downstream of the cylinder on the wake
centerline.  The cylinder was heated so slightly that the buoyancy
term was unimportant and temperature acted as a passive scalar.
Temperature was measured as a function of time, and we use here the
standard Taylor hypothesis that surrogates time derivatives for space
derivatives.  In Fig. 1 we display $J_{2n}(R)/(2n\kappa)$ as a function 
of $(2\kappa)^{-1}J_2S_{2n}(R)/S_2(R)$ for $n$ varying from 2 to 6, 
and for various $R$
values in the inertial range. We see that all the points fall on a
line whose slope is unity to high accuracy, and whose intercept (in
log-log plot) is very closely zero. This good agreement is a
confirmation of the validity of the fusion rules. In addition, this
agreement lends support to the {\em assumption} that $\eta$ is $n$ and
$R$ independent. It should be stressed that individual tests at
various values of $n$ as a function of $R$ corroborate the same
conclusion, i.e. Eq.(\ref{almostfinal}) is supported by the
experimental data with $C_{2n}$ being near unity.  The most sensitive
test of the alleged constancy of the coefficients $C_{2n}$ is obtained
by dividing $J_{2n}(R)$ by $n J_2S_{2n}(R)/S_2(R)$ for all the
available values $n$ and $R$. The result of such a test is shown in
Fig.2. We see that all the measured values of $C_{2n}$ are
concentrated within the interval $(0.75,1)$ for all separation within
the inertial interval. Considering the fact that the quantities
themselves vary in this region over five orders of magnitude, we
interpret this as a good indication for the independence of $C_{2n}$
of $R$ and $n$. The $R$ independence is very clear, and is a direct
test of the fusion rules.  The weak $n$ dependence seems to indicate
that $C_{2n}$ decreases slightly with $n$; this may arise from the
limited accuracy of the data. We are reluctant to make a strong claim
about the accuracy of 10'th or 12'th order structure functions.
%%%%%%%%%%%%%%%%%%%%%%%%%%%%%%%%%%%%%%%%%%%%%%%%%%%%%%%%%%%%%%%%%
\begin{figure}
%\epsfxsize=7.6truecm
%\epsfbox{fig2.eps}
\caption{ A detailed test of the coefficient $C_{2n}$, see text for
details. The symbols are the same as in Fig.1.  The small systematic
decrease of $C_{2n}$ with $n$ may be due to insufficient accuracy at
the tails of the probability distribution which become more important 
at large values of $n$.}
\label{Fig2}
\end{figure}
%%%%%%%%%%%%%%%%%%%%%%%%%%%%%%%%%%%%%%%%%%%%%%%%%%%%%%%%%%%%%%%%%%
Let us accept for now the evidence that the coefficients $C_{2n}$ in
Eq.(\ref{almostfinal}) are $n$-independent, and look for a way to
understand it.  Note that $J_{2n}(R)$ can be exactly written in terms
of conditional averages in the form
\begin{eqnarray}
J_{2n}(R) &=& -4n\kappa\int d \delta 
T(\B.r,\B.r+{\bbox R}) P[\delta T(\B.r,\B.r+{\bbox R})]
\label{cond}\\ &\times&[\delta T(\B.r,\B.r+{\bbox R})]^{2n-1} 
\left<\nabla^2 T(\B.r) |\delta T(\B.r,\B.r+{\bbox R})\right>
 \ , \nonumber
\end{eqnarray}
where $\delta T(\B.r,\B.r+{\bbox R}) \equiv T({\bbox r}+{\bbox R}) - T({\bbox
  r})$. We see that the conditional average $\left<\nabla^2 T(\B.r)
  |\delta T(\B.r,\B.r+{\bbox R})]\right>$ appears as a natural object
that needs to be determined. If we make the assumption that the
conditional average, which in general is a function of the two
variables $R$ and $\delta T$ is factorizable as a function of $R$
times a function of $\delta T$, then the only possible such form is
\FL
\begin{equation}
-4\kappa\left<\nabla^2 T(\B.r) |\delta T(\B.r,\B.r\!+\!\B.R)\right>
={J_2\over S_2(R)} \delta T(\B.r,\B.r\!+\!\B.R).
\label{linear}
\end{equation}
With this form in (\ref{cond}) we regain the RHS of
(\ref{almostfinal}), but since the conditional average cannot be a
function of $n$ the coefficient $C_{2n}$ {\em must} be
$n$-independent.  We note that Kraichnan conjectured that the
conditional average is {\em linear} in $\delta T(\B.r,\B.r+{\bbox R})$
in the context of the Kraichnan model, and numerical simulations
supporting this conjecture were presented\cite{95KYC}. Moreover,
linearity approximations on conditional average of the form $\langle
\nabla^2 X | X \rangle$ and relation like (\ref{almostfinal}) have
also been studied earlier by Ching\cite{93C} and by Pope and
Ching\cite{93PC}.  In this Letter we propose that the linearity of the
conditional average in $\delta T(\B.r,\B.r+{\bbox R})$ is a general
property of a wider variety of turbulent advection problems. From our
discussion, it is clear that other results on conditional statistics
can be derived in a similar fashion.
%%%%%%%%%%%%%%%%%%%%%%%%%%%%%%%%%%%%%%%%%%%%%%%%%%%%%%%%%%%%%%%%%%%
\begin{figure}
%\epsfxsize=8truecm
%\epsfbox{fig3.eps}
\caption{The conditional average in Eq.(16) as measured from the Yale
data [1] normalized by the measured value of $J_2/S_2(R)$ as a
function of $\delta T({\bbox r},{\bbox r}+{\bbox R})$ for three different
values of $R$. The different $R$ values are designated by triangles, 
squares and circles respectively.
}\label{Fig3}
\end{figure}
%%%%%%%%%%%%%%%%%%%%%%%%%%%%%%%%%%%%%%%%%%%%%%%%%%%%%%%%%%%%%%%%%%%

Before we proceed to the implications of (\ref{linear}) we present the
experimental evidence for its validity. In Fig.3 we present results
from the same data set that was used above.  We show the conditional
average as a function of $\delta T(\B.r,\B.r+{\bbox R})$ for various
values of $R$. The line passing through the data points is not a fit,
but rather the line required by Eq.(\ref{linear}). We note that points
belonging to different values of $R$ fall on the same line, indicating
that indeed the conditional average is a function of $\delta
T(\B.r,\B.r+{\bbox R})$ times a function of $R$, and that we identified
correctly the function of $R$ as $J_2/ S_2(R)$. To test the generality
of this result we analyzed a second data set from the convective hard
turbulence regime of the well documented Chicago experiment
\cite{87HCL,89SWL}.  The experiment was performed in a cylindrical box
of helium gas heated from below, and the Rayleigh number can be as
high as $10^{15}$. The box has a diameter of 20 cm and a height of 40
cm. The temperature at the center of the box was measured as a
function of time, and we use the same Taylor-hypothesis to analyze the
conditional average. The results are shown in Fig. 4.  Although we see
larger statistical scatter at the ends of the plot, the basic
assertion of linearity with the correct slope are confirmed.
%%%%%%%%%%%%%%%%%%%%%%%%%%%%%%%%%%%%%%%%%%%%%%%%%%%%%%%%%%%%%%%%%%%%%%
\begin{figure}
%\epsfxsize=7.9truecm
%\epsfbox{fig4.eps}
\caption{Same as Fig.3 but computed from the Chicago data, Refs. 2,3.}
\label{Fig4}
\end{figure}
%%%%%%%%%%%%%%%%%%%%%%%%%%%%%%%%%%%%%%%%%%%%%%%%%%%%%%%%%%%%%%%%%%%%%
As explained, the linearity of the conditional average in $\delta T$,
Eq.(\ref{linear}) was not derived from first principles. To stress the
theoretical interest in such a derivation we consider briefly another
form of $J_{2n}(R)$ that is obtained by moving around one of the
gradients in (\ref{J2n}). Up to a term that is negligible for $R$ in
the inertial range we can write
\begin{equation}
J_{2n}(R)=-4n(2n-1)\kappa\left<|\nabla T(\B.r)|^2[T({\bbox r}+{\bbox R})
 - T({\bbox r})]^{2n-2}\right>
\ . \label{J2n2}
\end{equation}
Accepting Eq.(\ref{almostfinal}) with $C_{2n}=1$ we can write 
\begin{eqnarray}
-4\kappa\left<|\nabla T(\B.r)|^2[T({\bbox r}+{\bbox R})
 - T({\bbox r})]^{2n-2}\right>\nonumber \\
= 
{J_2S_{2n}(R)/(2n-1) S_2(R)} \ . \label{ohboy} 
\end{eqnarray}
The LHS can be written, similarly to (\ref{cond}), in terms of the
conditional average $\left<|\bbox \nabla T(\B.r)|^2|\delta
  T(\B.r,\B.r+\B.R)\right>$. It is obvious however that now we {\em
  cannot} assume that this quantity factorizes into a function of
$\delta T$ times a function of $R$. If it did, the dependence on
$\delta T$ must have been $(\delta T)^2$ in order to give us
$S_{2n}(R)$ on the RHS of (\ref{ohboy}). But we can never obtain in
this way the explicit $1/(2n-1)$ factor. This underlines the fact that
the factorization in (\ref{linear}) is far from being obvious or
trivial. Currently we do not know the deep reason why the conditional
averages of $\nabla^2 T$ afford factorization. We pose this as an
important issue for further theoretical research.

Lastly, we comment on the implications of these findings for the
exponents $\zeta_n$.  As discussed above, if we know the functional
form of $J_{2n}$ and the coefficient, we can use the balance equation
(\ref{bal}) to compute the scaling exponents, provided that we know
the nonlinear term $D_{2n}$. In the context of the Kraichnan model the
latter is known exactly, and the balance equation leads to a quadratic
equation for the exponents $\zeta_n$, with the solution
\begin{equation}
\zeta_{2n} = \case{1}{2}[\zeta_2 -d +
\sqrt{(\zeta_2+d)^2+4d\zeta_2(n-1)}]\ . \label{zeta2n}
\end{equation}
These are the exponents that were conjectured by Kraichnan. We note
that these exponents are in disagreement with the calculations of
refs.\cite{95CFKL,95GK} which attempted to compute the exponents by
perturbative methods using as a small parameter either $2-\zeta_2$ or
the inverse dimension $1/d$. If it turns out indeed that
(\ref{zeta2n}) is the correct non-perturbative result, one needs to
carefully rethink the meaning of these perturbative calculations.

\noindent
{\bf Acknowledgments}. IP is grateful for the C.N. Yang Visiting
Professorship of the Chinese University of Hong Kong that made our
collaboration possible. We thank K.R. Sreenivasan and A. Libchaber for
their experimental data. This work was supported in part by the German
Israeli Foundation, the Minerva Center for Nonlinear Physics, and the
Naftali and Anna Backenroth-Bronicki Fund for Research in Chaos and
Complexity. The work of ESCC is also supported in part by the Hong
Kong Research Grants Council (Grant No. 458/95P).

\end{document}